\documentclass[%
aps,
reprint,
superscriptaddress,
noeprint,
amsmath,amssymb,
prb,
floatfix,
]{revtex4-2}

\usepackage{chemformula}
\usepackage[utf8]{inputenc}
\usepackage[T1]{fontenc}
\usepackage{graphicx}
\usepackage{epstopdf}
\usepackage{siunitx}
\usepackage{amsmath}
\usepackage{amssymb}
\usepackage[colorlinks=true,allcolors=blue]{hyperref}
\usepackage{soul}
\usepackage[normalem]{ulem}
\usepackage{lipsum}

\begin{document}

\title{Structural and electronic properties of realistic two-dimensional amorphous topological insulators}

\author{Bruno Focassio}\email{b.focassio@ufabc.edu.br}
\affiliation{Federal University of ABC (UFABC), 09210-580 Santo André , São Paulo, Brazil}
\affiliation{Brazilian Nanotechnology National Laboratory (LNNano), CNPEM, 13083-970 Campinas, São Paulo, Brazil}

\author{Gabriel R. Schleder}
\affiliation{Federal University of ABC (UFABC), 09210-580 Santo André , São Paulo, Brazil}
\affiliation{Brazilian Nanotechnology National Laboratory (LNNano), CNPEM, 13083-970 Campinas, São Paulo, Brazil}

\author{Marcio Costa}
\affiliation{Instituto de F\'{\i}sica, Universidade Federal Fluminense, 24210-346 Niterói, Rio de Janeiro, Brazil}
\affiliation{Brazilian Nanotechnology National Laboratory (LNNano), CNPEM, 13083-970 Campinas, São Paulo, Brazil}

\author{Adalberto Fazzio}\email{adalberto.fazzio@lnnano.cnpem.br}
\affiliation{Brazilian Nanotechnology National Laboratory (LNNano), CNPEM, 13083-970 Campinas, São Paulo, Brazil}
\affiliation{Federal University of ABC (UFABC), 09210-580 Santo André , São Paulo, Brazil}

\author{Caio Lewenkopf}\email{caio_lewenkopf@id.uff.br}
\affiliation{Instituto de F\'{\i}sica, Universidade Federal Fluminense, 24210-346 Niterói, Rio de Janeiro, Brazil}

\date{\today}

\begin{abstract}
We investigate the structure and electronic spectra properties of two-dimensional amorphous bismuthene structures and show that these systems are topological insulators. 
We employ a realistic modeling of amorphous geometries together with density functional theory for electronic structure calculations. 
We investigate the system topological properties throughout the amorphization process and find that the robustness of the topological phase is associated with the spin-orbit coupling strength and size of the pristine topological gap. 
Using recursive non-equilibrium Green's function, we study the electronic transport properties of nanoribbons devices with lengths comparable to experimentally synthesized materials. 
We find a $2e^2/h$ conductance plateau within the topological gap and an onset of Anderson localization at the trivial insulator phase.
\end{abstract}

\maketitle 

\section{Introduction}

Topological phases of matter have gained great and increasing interest due to their outstanding properties and prospects of applications in spintronics, low energy loss devices, and quantum computing \cite{Moore2010a,Giustino2020}. Standard theoretical approaches for the search and classification of topological properties in materials are based on the analysis of the system symmetry and on topological invariants that rely on translational symmetry \cite{Bansil2016,HansanKane_colloquium,qshi_graphene_KaneMele,z2_KaneMele.95.146802,Fu2007a,Vergniory2019,Zhang2019,Tang2019}. 
However, the statement that topological properties are robust against disorder, one of the strongest predictions of the theory, clearly indicates that translational invariance is not a requirement. Nonetheless, the demonstration that non-crystalline systems can host non-trivial topology \cite{Agarwala2017} came as a surprise and opened interesting new investigation paths in the field.

Since \citet{Agarwala2017} have shown that a random lattice tight-binding model can display Chern insulator properties, several other non-crystalline lattice models with topological properties have been proposed \cite{Xiao2017,Poyhonen2018,Bourne2018,Agarwala2020,Yang2019,Mukati2020,Sahu2019,Varjas2019}. This kind of investigation also includes an experimental realization of a two-dimensional non-crystalline system with topological properties, namely, a system of coupled gyroscopes that hosts topological chiral edge modes \cite{Mitchell2018}.

The classification of topological phases in non-periodic systems also offers new challenges since the standard topological insulator invariants are not applicable \cite{Wright2014,Buchner2014,Marsal2020}. 
Instead, the strategies employed so far comprehend real-space topological markers \cite{Agarwala2017,Xiao2017,Mitchell2018, Poyhonen2018,Bourne2018,Agarwala2020}, response to external fields \cite{Mukati2020}, modified versions of the symmetry indicators \cite{Marsal2020}, Bott invariants \cite{Loring2010,Agarwala2017,Huang2018a,Huang2018b,Huang2019,Ni2020}, and others~\cite{Varjas2019,Huang2020a}. 

None of these studies takes into account the fundamental properties of amorphous materials \cite{Zallen_book}. For instance, unlike random lattices, amorphous systems show short-range order. To address real materials such properties need to be properly taken into account. Overcoming the realistic modeling challenges is essential for the systematic design and discovery of new amorphous topological materials.

Amorphous materials recently entered the hall of synthesized topological insulators (TIs). The crystalline phase of Bi$_2$Se$_3$ is a known 3D TI \cite{Zhang2009}, however, experimental evidence supports the existence of a surface Dirac cone with helical spin-texture also in its amorphous phase \cite{Corbae2019}. Such features are a hallmark of the quantum spin Hall (QSH) \cite{qshi_graphene_KaneMele,z2_KaneMele.95.146802} state, a topological phase protected by TRS. 
Two-dimensional trivial amorphous systems have also been synthesized, for instance, monolayer free-standing amorphous carbon \cite{Toh2020} revealed an atomic arrangement with a wide distribution of both bond lengths and angles, resembling more the crystallite model \cite{Wright2014} than the random network \cite{Zachariasen1932} for amorphous structures. Although metallic, this system allows us to aim for the realization of 2D amorphous topological phases with a similar structure.

There are several 2D topological materials \cite{Marrazo2019,TIs_ML,MeraAcosta2016a} that are candidates to serve as platforms for amorphous topological insulators. 
We choose to study the properties of amorphous bismuthene, since it has been already synthesized in the pristine form and due to its remarkable properties. The experimental band gap of Bi supported by SiC(0001) surface is \SI{0.67}{\electronvolt} and hosts a topologically non-trivial band structure with one of the largest reported topological band gaps \cite{Reis2017_Bi_on_SiC}. The robustness of flat bismuthene is such that it withstands $\sim 17\%$ of vacancy concentration while retaining its topological features, this threshold depends on the energy gap and spin-orbit coupling (SOC) strength \cite{Ni2020,pezo2020disorder}.

In previous work \cite{Costa2019}, some of us addressed the challenge of realistic modeling by first obtaining a material-specific amorphous topological insulator through the amorphization of flat bismuthene. However computational limitations restricted the analysis to small system sizes and amorphization steps, raising questions about robustness of the reported amorphous phase. This paper vastly expands the analysis of Ref. \cite{Costa2019}, confirming its main conclusion and reporting new findings.

Here, we systematically investigate the structural, electronic, and transport properties of different samples of amorphous topological insulators. Using density functional theory (DFT) \cite{dft1964,dft1965,review_dft_to_ml} calculations combined with an amorphization scheme we obtain bismuthene amorphous structures \cite{bond_flip_method,Costa2019}. We study their topological properties by calculating the topological invariants. We show that amorphization tends to suppress the band gap, but does not close it. We consider a two-terminal nanoribbon geometry and using non-equilibrium Green's functions (NEGF) \cite{transport_1_Caroli1971,trace_formula,transport_datta_book,transport_nardelli} combined with DFT we calculate the system linear conductance. We show that the edge states are robust even at realistic lengths, up to \SI{0.32}{\micro\metre}, with a conductance plateau of $2e^2/h$ for energies within the topological gap. Also, we consistently characterize the amorphization as a transformation that maps the pristine system into the amorphous one with the preservation of the QSH state linked to the SOC strength and size of the pristine topological band gap. 

This paper is organized as follows. In Sec.~\ref{sec:methods} we discuss the procedure used to generate the amorphous lattices. Additionally, we briefly show the standard methods we employ for the electronic structure and transport calculations. In Sec.~\ref{sec:results} we present our main results, namely, the topological characterization and the study of the structural, electronic, and transport properties of different realizations of amorphous bismuthene. 
We also show how to modify and control the material properties by breaking time-reversal symmetry. Our findings are summarized and discussed in Sec.~\ref{sec:conclusions}.

\section{Methods}
\label{sec:methods}

There is a vast literature on theoretical modeling of amorphous systems \cite{Zallen_book}. 
These works put forward different amorphization schemes and show that lattice sizes $ 10^2 \cdots 10^3$ atoms with periodic boundary conditions are sufficient to describe the experimental data
\cite{Wooten1985, Tu1998}. The simulations in these studies can rely on well established empirical force models. This is not the case for 2D bismuthene. Here we perform a full DFT relaxation calculation to generate amorphous structures. From the computational point of view, this is the most severe bottleneck of our study. The details of our procedure are presented below.

\begin{figure}[!htb]
	\centering
	\includegraphics[width=\linewidth]{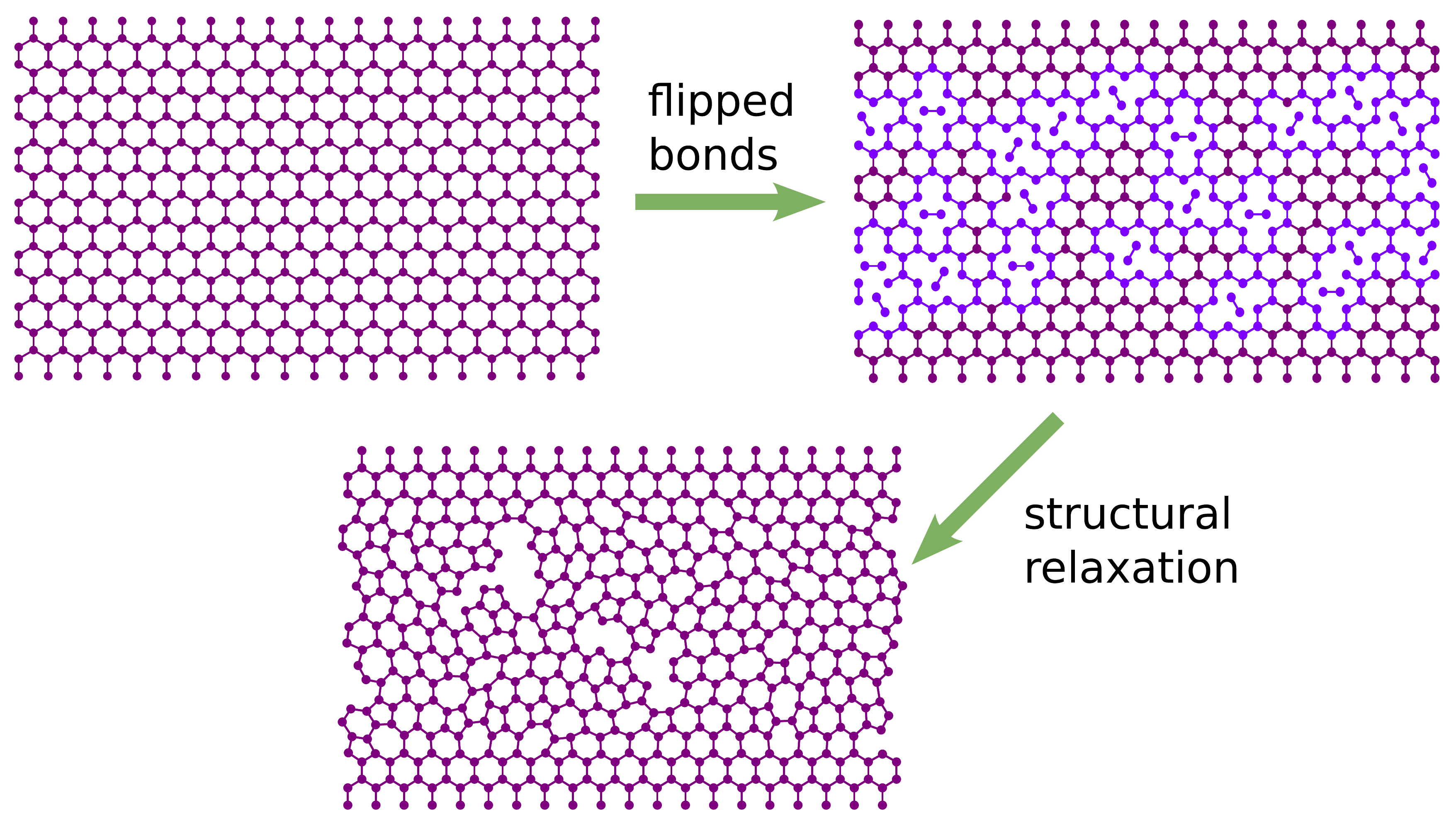}
	\caption{Representation of the bond flipping method \cite{bond_flip_method} for obtaining realistic amorphous geometries. The regions with flipped bonds are colored in blue.}
	\label{fig:amorphization_process}
\end{figure}

We create our amorphous geometries using the bond-flipping method \cite{bond_flip_method} together with density functional theory (DFT) \cite{dft1964,dft1965,review_dft_to_ml} for structure relaxation. Figure \ref{fig:amorphization_process} depicts the adopted procedure. \citet{Hsu2015b} demonstrated that hydrogen half-functionalized flat bismuthene (H-bismuthene) with SiC(0001) lattice parameter (\SI{5.35}{\angstrom}) essentially reproduces the same electronic structure without the computational cost of explicitly including the SiC surface. We start with a pristine flat bismuthene structure with 560 Bi and 560 H atoms. Within this geometry, the region to be transformed comprehends 400 Bi atoms and 400 H atoms, and we left pristine regions at each side so the following transport calculations can be performed. The number of bonds to be flipped is determined to result in a stable deviation of ring size distribution with the number of flipped bonds \cite{TopologicalInvestigationofTwoDimensionalAmorphousMaterials, amorphous_structure_silica_bilayers_Ru,Toh2020}. For the generated geometries this is achieved with 27 flipped bonds. At each amorphization step, we randomly chose and flip the bonds inside this central region. After flipping the bonds, we perform a geometry relaxation using the Vienna \textit{ab initio} Simulation Package (VASP) \cite{vasp1,vasp2} until the total force on atoms was less than \SI{E-2}{\electronvolt\per\angstrom}. The relaxations use the generalized gradient approximation (GGA) with the Perdew-Burke-Ernzerhof (PBE) ~\cite{pbe} functional for the electronic exchange and correlation interactions, and the projector augmented wave (PAW)~\cite{paw} method for ionic core potentials.

For the optimized amorphous geometries, we calculate the electronic structure using a local basis set. The local basis is needed to obtain converged \textit{ab initio} Hamiltonians and ovelap matrices, which are the main ingredients for transport properties calculations. The electronic structure is obtained using the SIESTA code \cite{siesta_code} with on-site self-consistent spin-orbit coupling (SOC)\cite{siesta_soc_1,siesta_soc_2}, fully relativistic norm-conversing pseudopotentials \cite{tm_pseudopotentials}, Perdew-Burke-Ernzerhof (PBE) exchange and correlation functional \cite{gga,pbe}, real space energy grid cut-off of 350 Ry, and an optimized single-$\zeta$ (SZ) basis set. The k-point density is set to $30\;/$\si{\per\angstrom} in the transport direction. We add \SI{15}{\angstrom} of vacuum space to avoid spurious interactions between periodic images in non-periodic directions. The analysis of results is partially aided by the sisl code \cite{zerothi_sisl}.

\begin{figure}
	\centering
	\includegraphics[width=\linewidth]{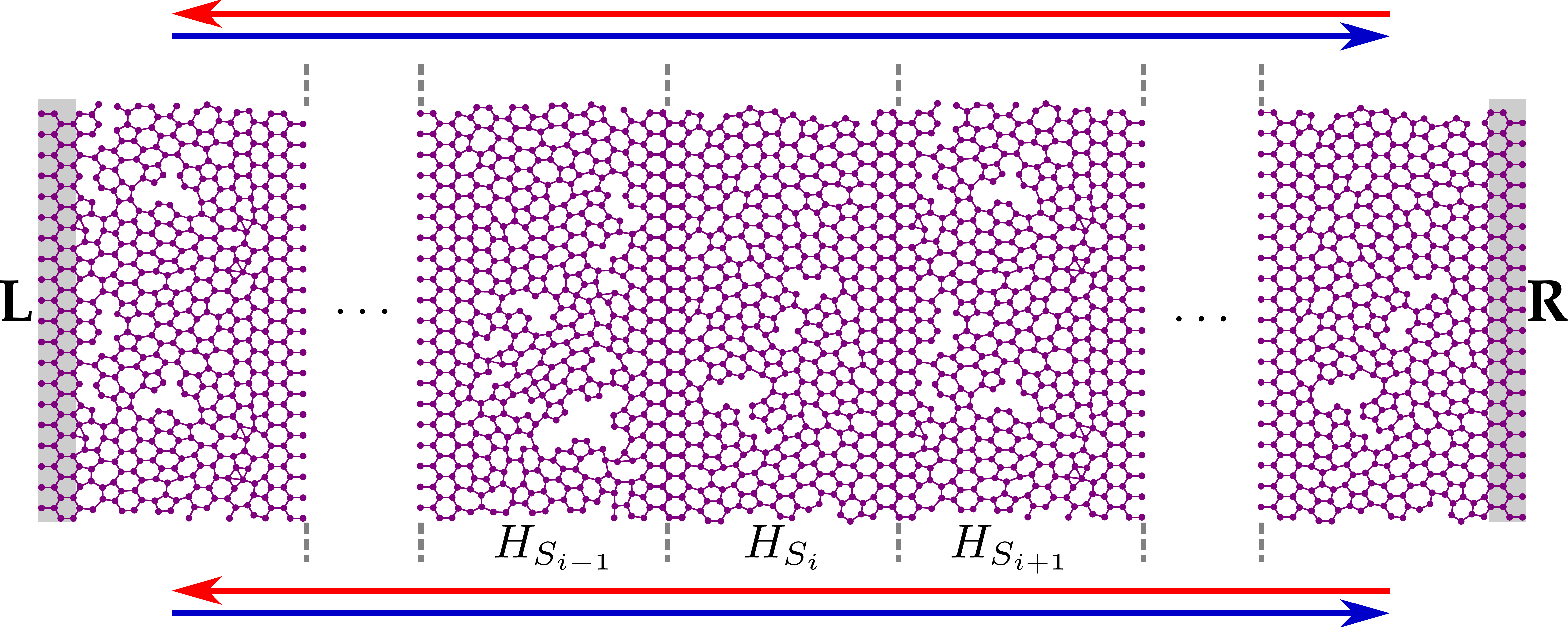}
	\caption{Two-terminal geometry considered in the NEGF calculations. The scattering region (S) is connected to left (L) and right (R) semi-infinite pristine bismuthene leads and the transport direction of the helical edge states is depicted by the red and blue arrows. The scattering region is partitioned into building blocks tailored for the decimation technique.}
	\label{fig:device}
\end{figure}

Using the real space Hamiltonians from SIESTA calculations, we implement the non-equillibrium Green's function (NEGF) \cite{transport_1_Caroli1971,trace_formula,transport_datta_book,transport_nardelli} method to obtain the system transport properties. For this, we consider a two-terminal setup with a scattering region (S) connected to leads in thermal and electrochemical equilibrium with left (L) and right (R) electron reservoirs as displayed in Fig.~\ref{fig:device}(a). The left (L) and right (R) leads are modeled as semi-infinite armchair flat bismuthene nanoribbons.

The system Hamiltonian is written in the spin resolved local basis as a block matrix,

\begin{equation}
\mathbf{H} = \begin{pmatrix}
\mathbf{H}_L & \mathbf{h}_{LS} & 0  \\
\mathbf{h}^\dagger_{LS} & \mathbf{H}_S & \mathbf{h}_{SR} \\
0 & \mathbf{h}_{SR}^\dagger & \mathbf{H}_R \\
\end{pmatrix},\label{eq:hamiltonian_two_lead_setup}
\end{equation}

\noindent where $\mathbf{H}_{L/R}$ is the Hamiltonian of the semi-infinite left (right) lead, $\mathbf{h}_{LS/SR}$ stands for the coupling matrix elements between the left (right) lead and the scattering region (S), and $\mathbf{H}_S$ is the scattering region Hamiltonian.

As standard \cite{transport_datta_book,transport_nardelli}, in the limit of small bias, the retarded Green's function reads

\begin{equation}
\mathbf{G}_S(E) = (E^{+}\mathbf{S} - \mathbf{H}_S - \mathbf{\Sigma}_L - \mathbf{\Sigma}_R)^{-1}, \label{eq:scatt_greensfunct}
\end{equation}

\noindent where $E^+ = \lim_{\delta \to 0^+} E+i\delta$, $\mathbf{S}$ is the local orbitals overlap matrix and $\mathbf{\Sigma}_{L/R}$ are the leads embedding self-energies given by $\mathbf{\Sigma}_{L} = \mathbf{h}_{LS}^{\dagger}\mathbf{g}_{L}\mathbf{h}_{LS}$ and $\mathbf{\Sigma}_{R} = \mathbf{h}_{SR}\mathbf{g}_{R}\mathbf{h}_{SR}^{\dagger}$. The leads surface Green's function ($\mathbf{g}_{L/R}$) is calculated following the Refs. \cite{transport_decimation_sanvito,Rocha2006}.

Finally, the system linear conductance is given by the Landauer formula \cite{transport_datta_book}, that is 

\begin{equation}
\mathcal{G} = \frac{e^2}{h} \int \text{d}E \left(- \frac{\partial f}{\partial E} \right)\mathcal{T}(E)   \label{eq:cond_temp}
\end{equation}

\noindent where $f$ is the Fermi-Dirac distribution and $\mathcal{T}$ is the transmission \cite{transport_1_Caroli1971,trace_formula} given by

\begin{equation}
\mathcal{T}(E) = \text{Tr}[\mathbf{\Gamma}_L \mathbf{G}_{S} \mathbf{\Gamma}_R \mathbf{G}_{S}^\dagger], \label{eq:transmission_as_trace}
\end{equation}

\noindent where $\mathbf{\Gamma}_{L/R} = i [\mathbf{\Sigma}_{L/R} - \mathbf{\Sigma}_{L/R}^{\dagger}]$ are the decay width matrices coupling the leads and the scattering region. At zero temperature, Eq. \eqref{eq:cond_temp} is reduced to $\mathcal{G} = (e^2 / h ) \mathcal{T}(E)$.

Using the scattering region Green's function we can also compute the electronic density of states (DOS), namely,

\begin{equation}
{\rm DOS(E)} = \frac{1}{\pi}\;{\rm Im}\;{\rm Tr} [ \mathbf{G}_S(E) ], \label{eq:negf_dos}
\end{equation}

The computation of the Green's function is a daunting task for a realistic size model system with a large basis of local orbitals $N_T$, since Eq. \eqref{eq:scatt_greensfunct} requires the inversion of a $N_T \times N_T$ matrix. For disorded systems, where one is interested in ensemble averages, the computational time becomes even more critical. This problem is mitigated by the recursive Green's function (RGF) methods that explore the fact that only a small fraction of the matrix elements of $\mathbf{G}_S(E)$ are necessary to compute $\mathcal{T}$, namely the ones that are connected to the $L/R$ leads.

The standard implementation of the RGF is to partition the scattering region into $N$-building blocks as show in Fig. \ref{fig:device}. Here, we consider building blocks that are sufficiently large to simulate amorphous domains (as discussed above). Such large size building blocks guarantees that only nearest neighbor partitions are coupled, allowing to write $\mathbf{H}_S$ as

\begin{equation}
\mathbf{H}_S = \begin{pmatrix}
\mathbf{H}_{S_1} & \mathbf{V}_{1,2} & 0 & \hdots & 0 \\
\mathbf{V}_{1,2}^\dagger & \mathbf{H}_{S_2} & \hdots & \hdots &\vdots \\
0 & \vdots & \ddots &\vdots &0 \\
\vdots & \hdots & \hdots &  \mathbf{H}_{S_{N-1}} & \mathbf{V}_{N-1,N} \\
0 & \hdots & 0 & \mathbf{V}_{N-1,N}^\dagger & \mathbf{H}_{S_N} \\
\end{pmatrix}\label{eq:scatt_hamiltonian}
\end{equation}

\noindent where $\mathbf{H}_{S_i}$ is the $i$th building block Hamiltonian and $\mathbf{V}_{i,j}$ represents its coupling to the $j$th block. This partition scheme is depicted in Fig. \ref{fig:device}. The $\mathbf{V}_{i,j}$'s are taken as constant and corresponding to the coupling of a pristine block to its neighbors, which is guaranteed by the buffer layers as seen in Fig. \ref{fig:device}.

The recursive method is used to eliminate by decimation the matrix elements of the local sites which are not coupled to the electrodes. We use the procedure put forward in Ref. \cite{transport_decimation_sanvito} and generalized to the case of $\mathbf{S} \neq \mathbf{I}$ in Ref. \cite{reily_2010_decimation_theory}. In this way, the computational cost is reduced by a factor of $N^2$ as compared with the full matrix inversion of Eq. \eqref{eq:scatt_greensfunct}.

The partition scheme also allow us to combine different sequences of amorphous supercells to build nanoribbons of arbitrary length. The strategy has been successfully applied to study the conductance of different disordered systems \cite{transport_decimation_sanvito,disorder_si_wires_markussen,Rocha2006,Lewenkopf2013a_review_recursiveGF,reily_2010_decimation_theory,reily_design_real_nanotube,disorder_graphene_spintronics,disorder_cnt_james}.

\section{Results}
\label{sec:results}

\begin{figure*}
	\centering
	\includegraphics[width=\linewidth]{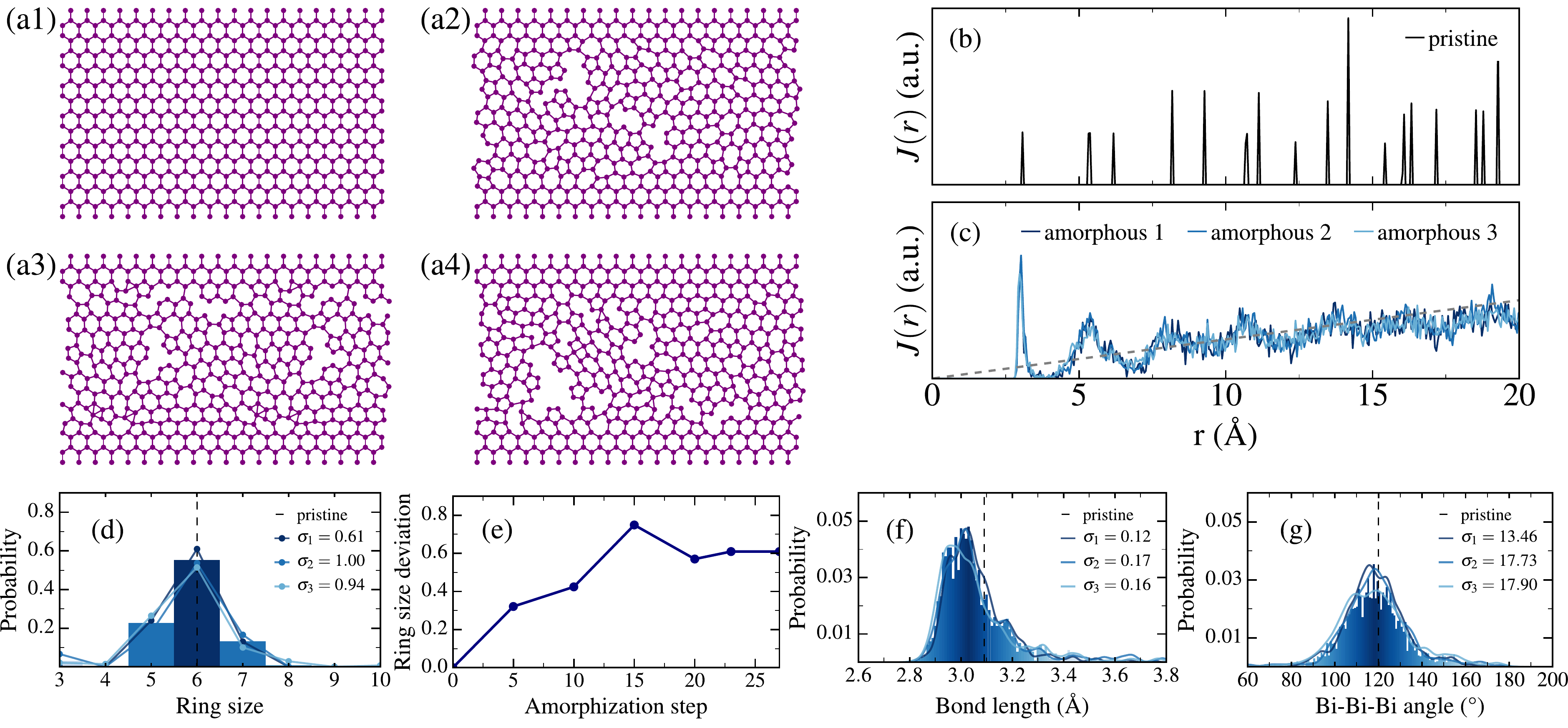}
	\caption{(a) H-bismuthene geometries, (a1) is the pristine system and (a2)--(a4) shows different amorphous realizations named 1, 2 and 3, respectively. Partial bismuth radial distribution function $J(r)$ for (b) pristine geometry and (c) the three amorphous geometries without the pristine leads. In panel (c) the dashed line marks the usual RDF behavior for 2D amorphous materials, namely, $J(r)\propto r$ for $r/a \gg 1$, with $a$ the lattice parameter. (d) Shows the ring size distribution. (e) Is the evolution of the ring size standard deviation in function of the amorphization step for the amorphous realization \#1, in panel (a2). (f) Shows the bond length distribution, and (g) Shows the Bi--Bi--Bi bond angle distribution. In panels (d),(f), and (g) the histogram shows the mean distribution between the different amorphous realizations and $\sigma$ is the standard deviation.}
	\label{fig:geom_pdf_fft}
\end{figure*}

\subsection{Amorphous structures}
\label{sec:amorphous_structures}

We start by generating three amorphous H-bismuthene structures by the aforementioned bond-flip method. The corresponding obtained amorphous geometries after 27 amorphization steps are shown in Fig. \ref{fig:geom_pdf_fft}(a). Each amorphous region contains 400 Bi and 400 H atoms, spanning a periodic lattice (without the vacuum layer) with \SI{107}{\angstrom} width and \SI{46.33}{\angstrom} length along the transport direction (see Fig. \ref{fig:device}(a)). One should note the addition of pristine buffer layers to the geometries so each system could be properly coupled along the transport direction for transport calculations. Each lead/buffer layer corresponds to a pristine armchair ribbon with 80 Bi and 80 H atoms with \SI{9.27}{\angstrom} length. To address the structural characterization and amorphization degree of these structures, we inspect the first neighbors connectivity, the radial distribution function (RDF), ring size statistics, bond length and bond angle distributions.

In Fig. \ref{fig:geom_pdf_fft}(a2)--(a4) we set the maximum bond length to $\SI{3.80}{\angstrom}$ which is $\approx 20\%$ larger than the pristine bond length ($\SI{3.09}{\angstrom}$). We observe that the amorphous bismuthene structures are significantly less connected than the pristine ones. We stress that all these metastable structures are obtained by \textit{ab initio} relaxation calculations. 
This result indicates that the standard approach of continuous random models to consider a fixed coordination number cannot be applied to H-bismuthene. In the next sections we discuss the consequences of the formation of holes in the electronic properties of the system.

Information regarding short and long range order is obtained through the radial distribution function (RDF) analysis \cite{Levine2011,Schleder2019_ligand_exchange}. We computed the partial Bi RDF of the amorphous regions via the histogram method implemented in the VMD software \cite{Humphrey1996_VMD}. 
Figure ~\ref{fig:geom_pdf_fft}(b) shows that the crystalline system yields sharp and well-defined peaks at the corresponding neighbor distances. In contrast, Fig.~\ref{fig:geom_pdf_fft}(c) shows that in the amorphous systems short-range order is characterized by two isolated peaks, the first at the first neighbors distance at $r \approx \SI{3.09}{\angstrom}$, which is close to the Bi--Bi pristine bond length and a second near $\SI{5.30}{\angstrom}$ corresponds to the second neighbor distance. As $r$ increases the peaks become broader and undefined as a result of the absence of long range order. Figure \ref{fig:geom_pdf_fft}(c) clearly shows that $J(r) \propto r$ for $r/a \gg 1$ as expected for 2D amorphous materials \cite{Zallen_book,Buchner2014,Costa2019}. We have also confirmed that the structure factor (not shown here) is isotropic, as expected. This analysis shows that the system sizes we study are large enough to describe amorphous systems.

In Fig. \ref{fig:geom_pdf_fft}(d) we present the ring size distribution for the amorphous geometries \cite{LeRoux2010}. The amorphization is quantified considering the standard deviation from hexagons. Each structure presents a different $\sigma$, though they all show a broad ring size distribution as expected for amorphous materials \cite{Buchner2014}. Moreover, for the amorphous realization \#1 (Fig. \ref{fig:geom_pdf_fft}(a2)), in Fig. \ref{fig:geom_pdf_fft}(e) we monitor the ring size deviation with the number of flipped bonds to find that the final deviation is stable for the last amorphization steps, revealing that no further structural modifications are obtained by continuing the amorphization process.

Figures \ref{fig:geom_pdf_fft}(f) and (g) show the bond length and bond angle distributions. We observe a contraction in bond length as the mean value for the amorphous structures is \SI{3.06}{\angstrom}, less than the pristine value, which results in the presence of non-connected regions inside each amorphous geometry. The mean Bi--Bi--Bi bond angle is still centered around the pristine value which is expected since the average coordination number is close to $2.7$ for all structures.

\subsection{Electronic structure}
\label{sec:electronic_structure}

Let us now calculate the electronic structure of the amorphous realizations and characterize their topological properties. We investigate the energy levels of pristine and amorphous systems, the presence of metallic edge states at their interface with vacuum, and compute their topological indices.

\begin{figure}
	\centering
	\includegraphics[width=\linewidth]{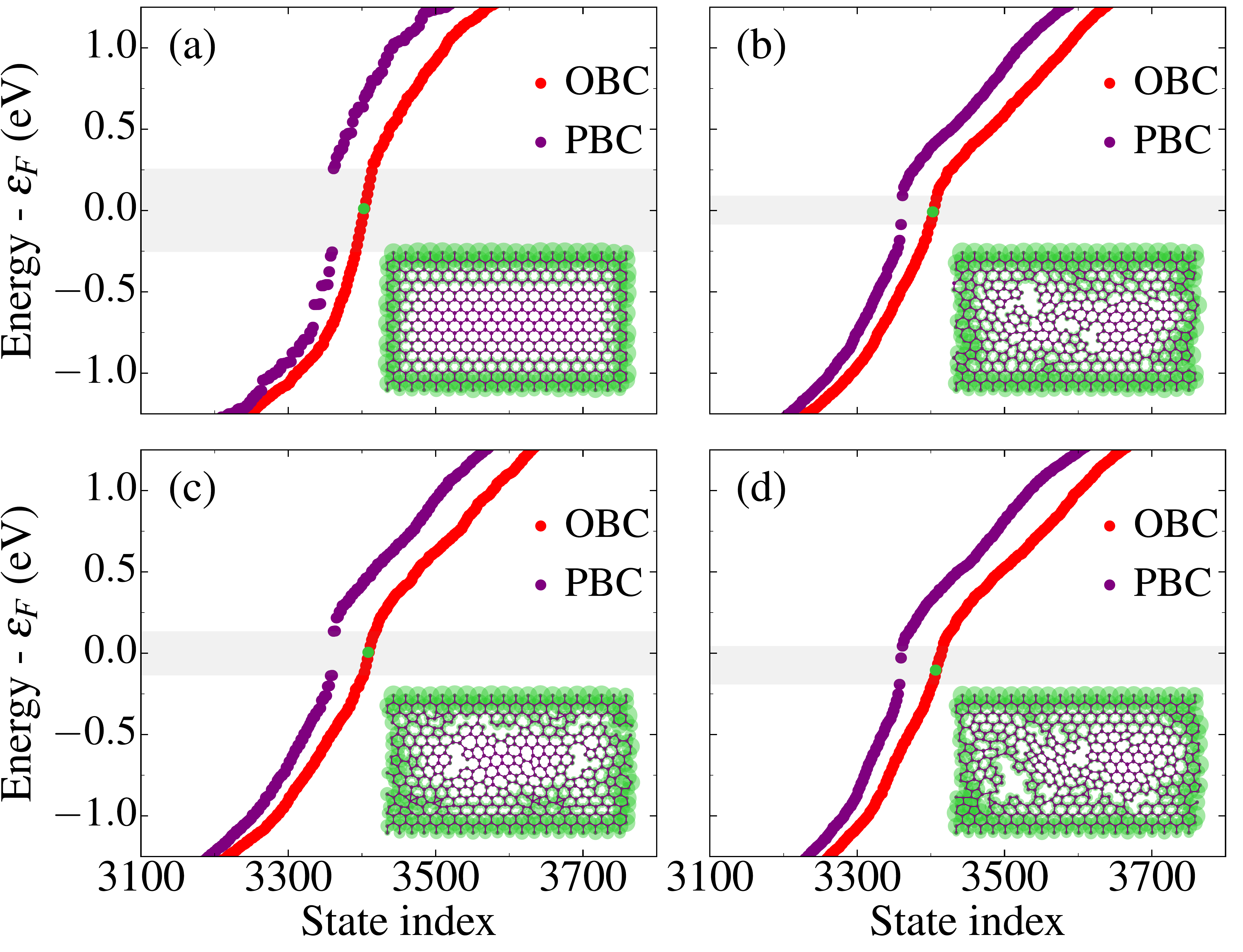}
	\caption{Energy eigenvalues \textit{versus} state index for periodic (PBC) and open boundary conditions (OBC). Panel (a) corresponds to the pristine system, and (b) to (d) to the different amorphous realizations, as indicated by the insets. The gray regions stand for the topological gaps. The insets show the wavefunction probability amplitude of the states marked in green.}
	\label{fig:energy_state_wfc}
\end{figure}

For the case of periodic boundary conditions (PBC), these systems are insulators, with the band gap marked by the gray region in Fig. \ref{fig:energy_state_wfc}(a)--(d). The band gap of crystalline H-bismuthene is \SI{0.51}{\electronvolt}, while the amorphous band gap ranges from \SI{0.27}{\electronvolt} to \SI{0.16}{\electronvolt}. The decrease in band gap is a result of amorphization due to the degree of structural disorder and is realization dependent. 
The amorphization scheme we adopt can be considered as a sequence of discrete transformations that map the pristine insulating system into an amorphous insulating phase without a band gap closure. 
The topological nature of these systems is confirmed by inspecting their topological invariants. The $\mathbb{Z}_2$ invariant is computed using the Wannier charge centers (WCCs) evolution method \cite{z2pack_1,z2pack_2} implemented in the Z2Pack code \cite{z2pack_code}. For the crystalline system, the non-trivial topology is characterized by $\mathbb{Z}_2 = 1$. We consider the three amorphous structures we have generated as supercells and obtain $\mathbb{Z}_2 = 1$. For non-crystalline systems, the topological nature can also be characterized by the spin Bott index ($B_s$) that give a measure consistent with the $\mathbb{Z}_2$ invariant for sufficiently large system sizes, as discussed in Refs.~\cite{Huang2018a,Huang2018b}. Such problem is intrinsically avoided in non-crystalline systems due to the extended size of the geometries used to accommodate the amorphous phase. 
We implemented the spin Bott index as in Refs.~\cite{Huang2018a,Huang2018b} using the full \textit{ab initio} Hamiltonians in local basis, and obtain $B_s = 1$ for the three amorphous systems. Both results are consistent and show that bismuthene keeps its pristine topological features in the amorphous state.

For open boundary conditions (OBC), the energy gap is filled. In the pristine case the states filling the topological gap are localized at the system edges. 
Conversely, for the amorphous systems, the distribution of these wavefunctions along the edges is significantly altered by the amorphization with respect to the crystalline case. However, we still observe strong edge localization with a variation along the edge, analogous to Ref. \cite{Costa2019} and other lattice models of amorphous TIs \cite{Agarwala2017,Mitchell2018,Xiao2017,Poyhonen2018,Huang2018a,Huang2018b,Marsal2020}. The insets in Fig. \ref{fig:energy_state_wfc} show the site-projected wavefunction of selected states.

We stress that, even though all considered structures have the same number of bonds flipped, the relaxation yields geometries with different connectivities, that directly affect the energy gap, as seen in Fig. \ref{fig:energy_state_wfc}. This effect is more drastic for the amorphous realization \#3 (see. Fig.~\ref{fig:geom_pdf_fft}(a4) Fig.~\ref{fig:energy_state_wfc}(d)) since bulk defect-like states appear inside the energy gap pinning the Fermi level. For this realization, the lowest occupied state is localized at the void region in the bulk, and close analysis shows that the real energy gap is 0.16 eV. As a result, the selected state shows edge localization even though is below the Fermi energy. 
A similar feature of has been reported in a theoretical study of vacancies and vacancy clusters in flat bismuthene \cite{Ni2020}. The robustness of the QSH state is proportional to the pristine energy gap, hence the material supports the non-trivial topological state until the defect concentration reaches a threshold value. This observation raises the question: \textit{Does the topological gap close as one considers systems with more bond flips?}

\begin{figure}
	\centering
	\includegraphics[width=\linewidth]{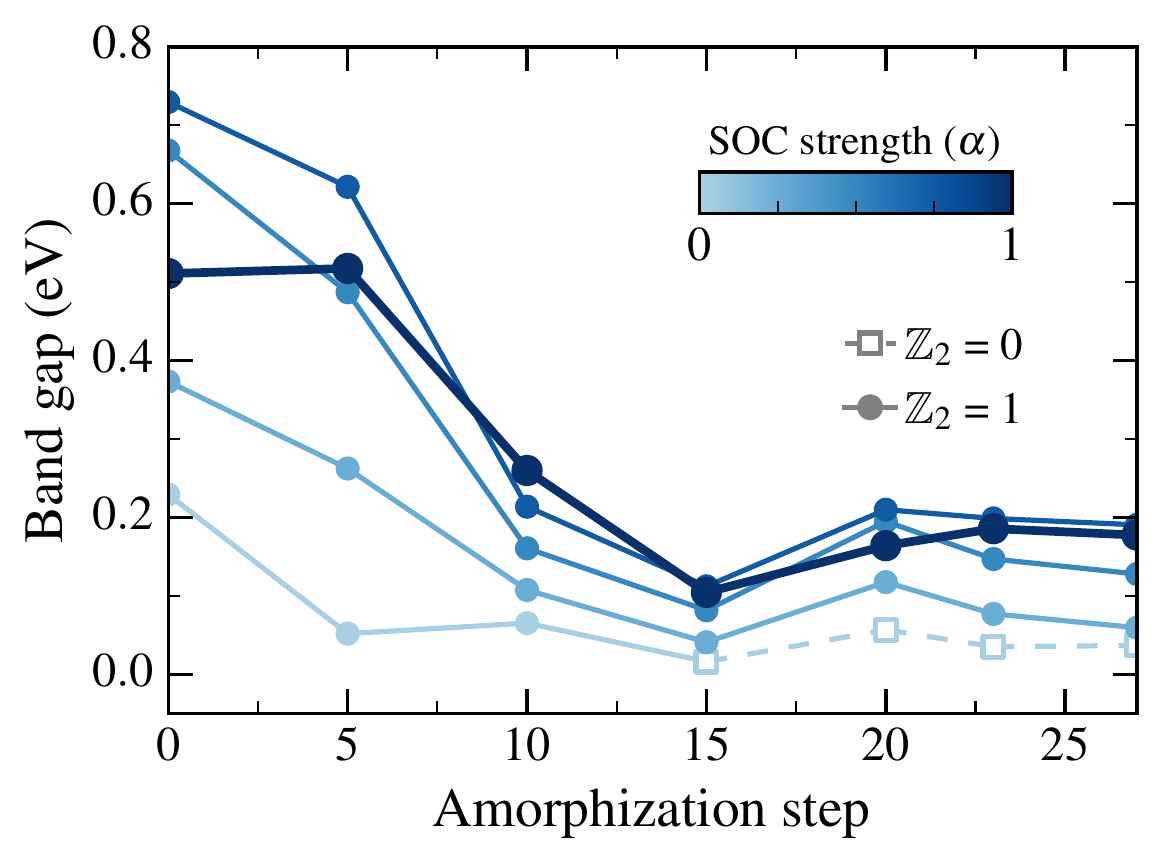}
	\caption{Evolution of the band gap as a function of the amorphization for the amorphous structure \#1 (Fig. \ref{fig:geom_pdf_fft}(a2)). The evolution is tracked for different spin-orbit coupling strengths ($\alpha$), namely, $\alpha = $ $0.01$, $0.25$, $0.50$, $0.75$ and $1.00$. For all colors, filled circular markers with solid lines correspond to $\mathbb{Z}_2 = 1$ and empty square markers with dashed lines correspond to $\mathbb{Z}_2 = 0$.
	}
	\label{fig:soc_strength}
\end{figure}

In all studied cases the energy gap shows an overall decrease as a function of the amorphization step, but for a realistic number of flipped bonds it never closes, yielding the topological amorphous insulators presented here. Figure \ref{fig:soc_strength} shows the gap as a function of the amorphization step for the amorphous realization \#1. The similar behavior holds for the three amorphous realizations with full SOC strength: The $\mathbb{Z}_2$ is kept invariant and there are no signs of a quantum phase transition. Furthermore, since the standard deviation of ring size with amorphization step shows a saturation for the last amorphization steps (see Fig. \ref{fig:geom_pdf_fft}(e)) indicating that no significant structural changes occur for more bond flips. Hence, we also expect that the electronic structure is stabilized.

Next, we study the dependence of the robustness of the topological gap on the spin-orbit interaction strength. For that, we analyze the evolution of the band gap by artificially quenching the SOC strength by a factor $\alpha$ ($0 < \alpha < 1$). First, regardless of the amorphization, for $\alpha = 0$ the system is metallic. The result in Fig. \ref{fig:soc_strength} is obtained for the amorphous realization in Fig. \ref{fig:geom_pdf_fft}(a2) (\#1). The pristine system (Fig. \ref{fig:soc_strength}, amorphization step 0), shows a surprising behavior: For small to intermediate SOC strengths, the band gap increased with $\alpha$, reaching its maximum value at $\alpha \approx 0.75$. By further increasing $\alpha$, the band gap decreases to the actual value of flat bismuthene at $\alpha = 1$. This finding suggests exploring the possibility of designing novel QSHIs with increased robustness by alloying with elements with smaller SOC strength than Bi, for instance, Sb. Figure \ref{fig:soc_strength} shows that for large amorphization steps, the band gap stabilize regardless of further geometry modifications. For intermediate steps the band gap decreases, reaching a minimum then increasing to its final value. This minimum is reached for 15 steps. For the QSH state to survive the amorphization the energy gap should remain open at this minimum, which is clearly dependent on the SOC strength. At this step, the energy gap converges to $0.11\;\rm eV$ with increasing SOC strength. For $\alpha = 0.01$, the band gap reaches a minimum of $0.02\;\rm eV$ indicating that a quantum phase transition may occur for this SOC strength before $15$ amorphization steps. This is further supported by calculating the $\mathbb{Z}_2$ invariant. For all SOC strengths larger than $0.01$, we obtain $\mathbb{Z}_2 = 1$ throughout the amorphization procedure. For the case of $\alpha = 0.01$ SOC strength, we find $\mathbb{Z}_2 = 1$ before the minimum energy gap is reached, while $\mathbb{Z}_2 = 0$ after the energy gap reopens, signaling a quantum phase transition from the topologically non-trivial to the trivial system. Therefore, the robustness to amorphization is system dependent, although the topology is maintained for systems with medium to strong SOC strengths.

\begin{figure}[htb]
	\centering
	\includegraphics[width=\linewidth]{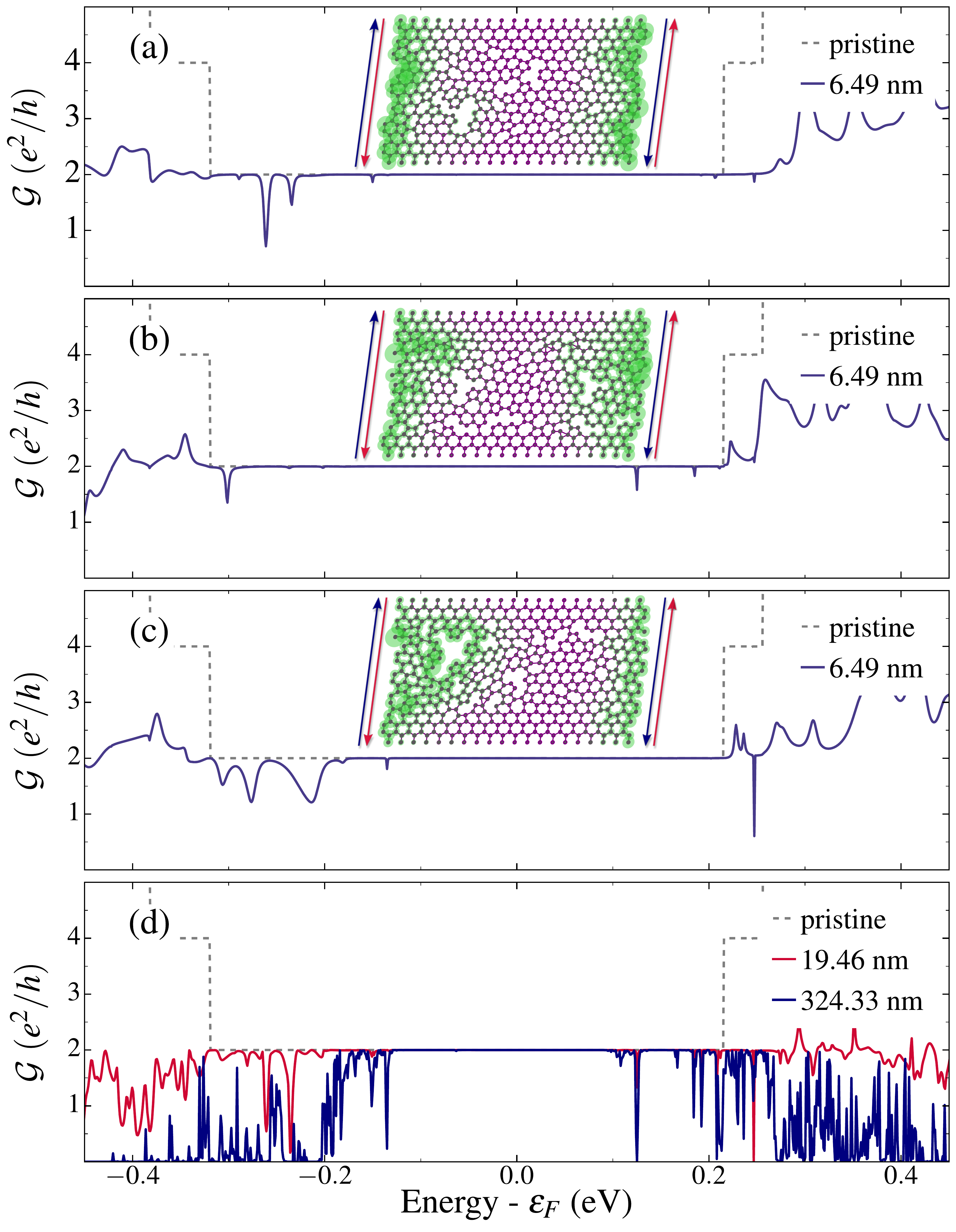}
	\caption{Electronic linear conductance $\mathcal{G}$ for different amorphous realizations and device lengths as a function of the energy. Panels (a) to (c) corresponds to the different amorphous realizations. Panel (d) corresponds to a scattering region composed of randomly selected sequences of amorphous realizations. The insets shows the site-projected wavefunction probability amplitude for the states with energy $E=\varepsilon_F$. The arrows indicate the transport direction.}
	\label{fig:trc_length}
\end{figure}

\subsection{Electronic transport}
\label{sec:transport}

Here, we study the effect of the non-trivial topology in the electronic conductance of amorphous bismuthene nanoribbons using the Green's function approach described in Sec. \ref{sec:methods}. The crystalline nanoribbons have $\mathcal{G}_0$ ($2e^2/h$) conductance inside the topological gap, which is a hallmark of the metallic helical edge states of QSHIs. As the energy is moved away from the topological gap, the conductance is increased in integer steps of $\mathcal{G}_0$, due to the opening of modes corresponding to the bulk band structure of the nanoribbon geometry. See dashed lines in Fig. ~\ref{fig:trc_length}.

We use the three amorphous bismuthene structures obtained as discussed above to build amorphous nanoribbons. For single building blocks, the latter have a scattering region of $6.49\rm\;nm$ in length. Figures ~\ref{fig:trc_length}(a), ~\ref{fig:trc_length}(b) and ~\ref{fig:trc_length}(c) show the corresponding calculated conductance as a function of the energy. The $\mathcal{G}_0$ conductance inside the topological gap is a fingerprint of the topological edge states' robustness and protection against back-scattering. For trivial states, outside the topological gap, the conductance is strongly suppressed with respect to the crystalline one, as expected. These results are in agreement with the results of Ref. \cite{Costa2019} for amorphous H-bismuthene and defective topological systems, including stanene \cite{Tiwari2019a_transport_vacancies_stanene} and 2D Na$_3$Bi \cite{Focassio2020}. The insets in Fig.~\ref{fig:trc_length} show the amorphous ribbons with site-projected wavefunction probability amplitude at $E=\varepsilon_F$. These plots show that, despite being strongly affected by amorphization, the edge states preserve their key features, namely they are extended and states at opposite edges of the ribbon do not hybridize.

We now study the conductance for realistic device lengths by considering the three amorphous structures obtained above as building blocks for longer nanoribbons and using the recursive method to compute the transmission. First we built a system consisting of 3 building blocks in a row. The resulting scattering region is $19.46\;\rm nm$ long. The corresponding conductance, shown by the red line in Fig.~\ref{fig:trc_length}(d), is qualitatively similar to the ones of Figs.~\ref{fig:trc_length}(a) to (c). Next, we randomly combine the amorphous structures to form a sequence of 50 building blocks, that gives a nanoribbon of $324.33\;\rm nm$ in length. The calculated conductance, blue line in Fig.~\ref{fig:trc_length}(d), shows $\mathcal{G}=\mathcal{G}_0$ for $-0.13\;{\rm eV} < E < 0.10\;{\rm eV}$, confirming the robustness of the QSH phase against amorphization, in line with the results presented in Sec. \ref{sec:electronic_structure}. For energies outside this range but still within the bulk topological gap, the conductance is suppressed at few narrow energy intervals. The latter become more dense as the energy approaches the bottom (or top) of the non-topological bands. For energies corresponding to trivial states, the conductance is strongly suppressed and the system becomes an insulator.

\begin{figure}[htb]
	\centering
	\includegraphics[width=\linewidth]{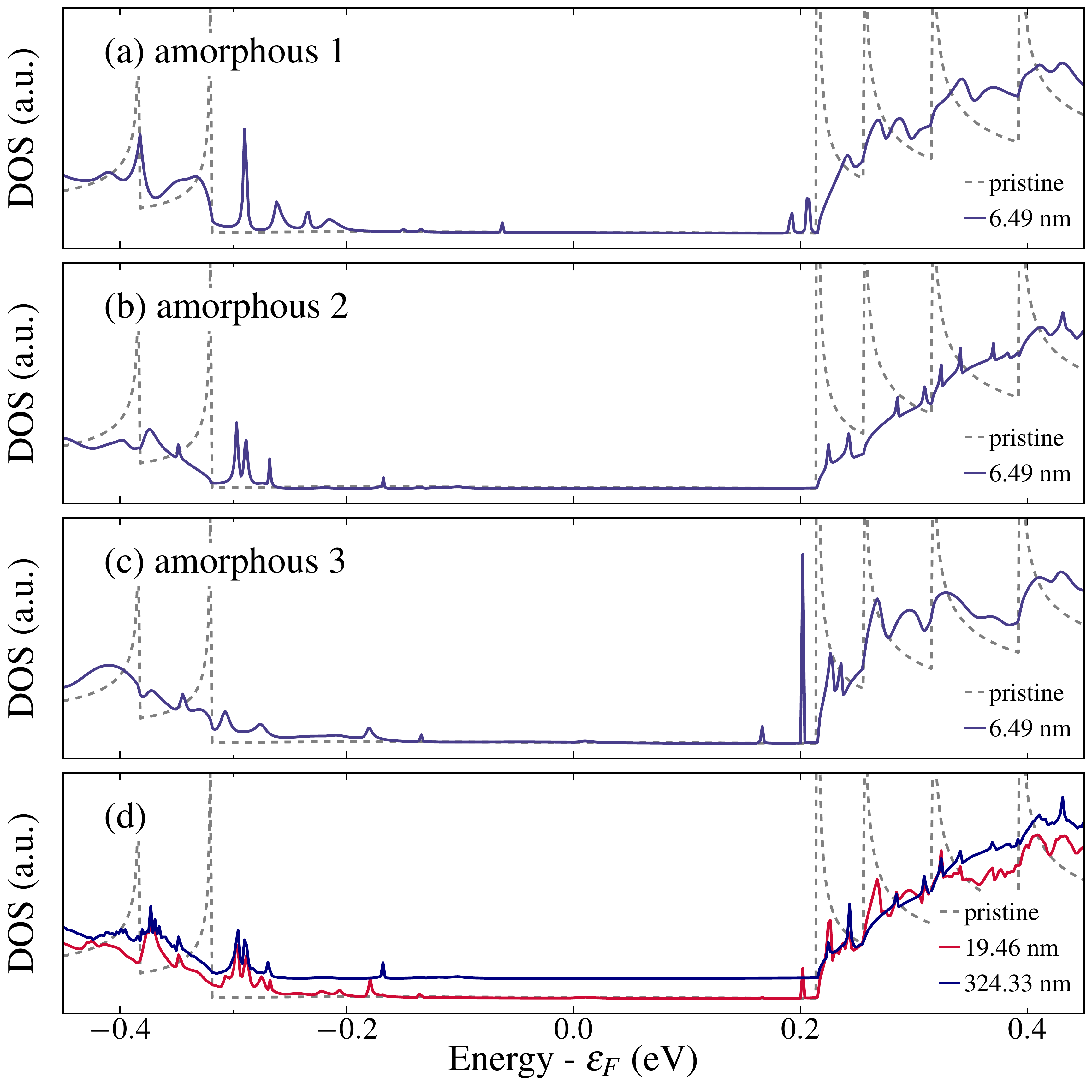}
	\caption{Density of states as a function of the energy. The nanoribbons are the same as in Fig. \ref{fig:trc_length}. The DOS in panel (d) are shifted in the $y$-axis for better visualization. The pristine nanoribbon is represented by the dashed line.}
	\label{fig:dos_amorfo}
\end{figure}

To further investigate this feature, we analyze the DOS of the systems addressed in Fig. \ref{fig:trc_length}. The dashed lines in Fig. \ref{fig:dos_amorfo} correspond to the DOS of a pristine bismuthene nanoribbon and serve to guide our discussion. The amorphization drives electronic states into the topological gap. When these states are isolated, they do not affect the electronic transport, an indication that they are localized. For energies close to the van Hove singularities but still in the topological gap, the density of states driven by amorphization increases and due to their overlap the conductance decreases, see Fig. \ref{fig:trc_length}. The situation is very different for energies outside the topological gap. Here, while the DOS increased significantly, the conductance if strongly suppressed and the system becomes an insulator. This behavior is characteristic of the onset of Anderson localization \cite{VanTuan2012,Lherbier2013}.

\subsection{Breaking time-reversal symmetry}
\label{sec:application}

\begin{figure}[htb]
	\centering
	\includegraphics[width=\linewidth]{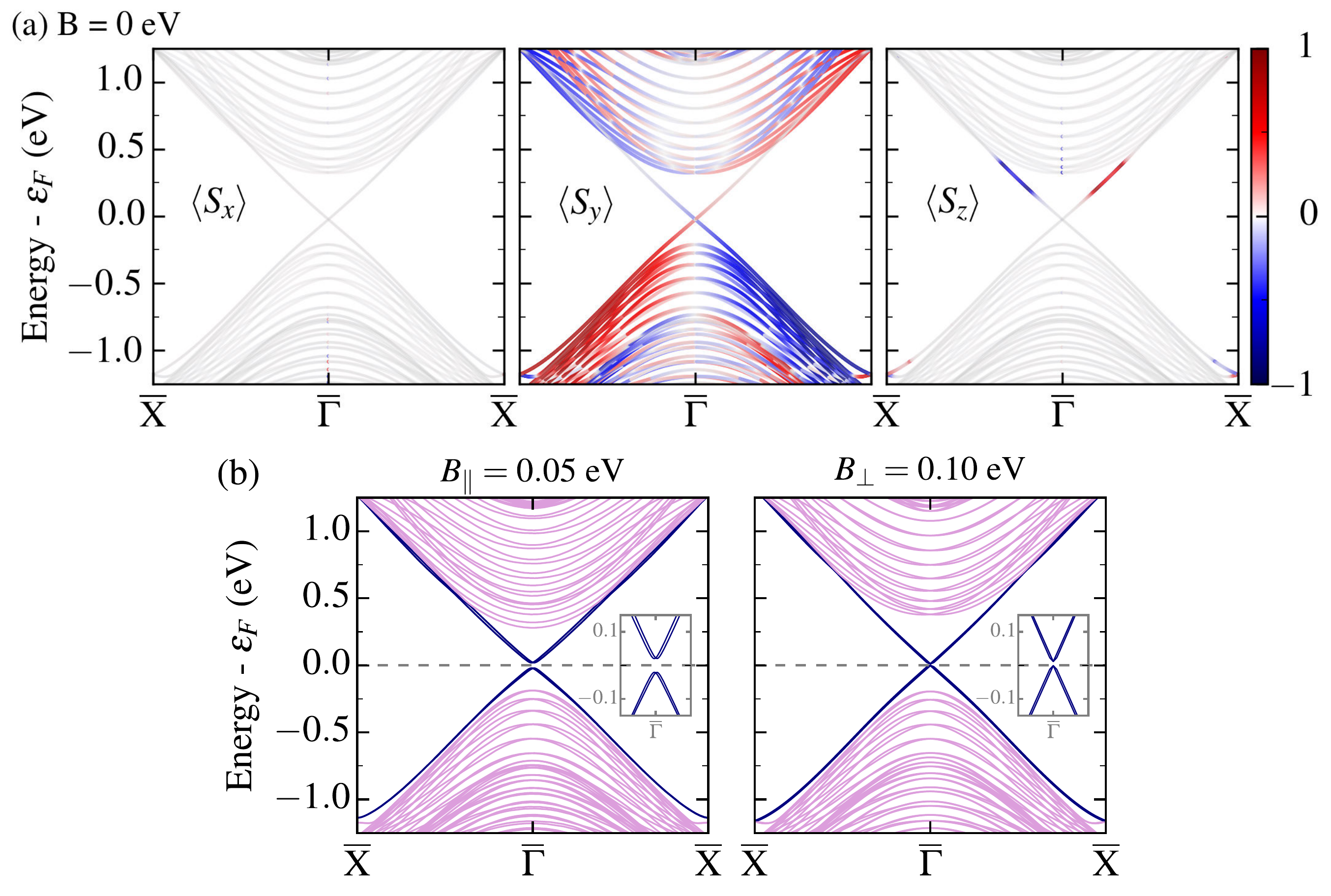}
	\caption{(a) Spin texture for armchair pristine H-bismuthene nanoribbon with $B = \SI{0.0}{\electronvolt}$. The color bar is in units of $\hbar/2$. (b) Armchair nanoribbon band structure for parallel ($B_\parallel$) and perpendicular ($B_\perp$) exchange field.}
	\label{fig:texture_field_bands}
\end{figure}

Let us now investigate possible mechanisms to manipulate the system conductance of the amorphous topological insulators protected by TRS. This can be accomplished by, for instance, adding to the Hamiltonian a simple exchange field expressed as an on-site term of the form $\delta H = \mathbf{B} \cdot {\boldsymbol \sigma}$. The computation of the perturbation term is done by calculating the expectation value of the Pauli matrices $\sigma_i$ $(i=x,y,z)$ on the local basis, which allow us to apply this field in different directions. This field opens an energy gap in the pristine nanoribbon, as seen in Fig.~\ref{fig:texture_field_bands}(b). The difference in gap opening is a result of the unbalanced spin-texture for the pristine nanoribbon. Figure \ref{fig:texture_field_bands}(a) shows the expectation values of the spin projection operator $\mathbf{S} = \hbar/2\;\boldsymbol \sigma$ for the pristine armchair nanorribbon. The spin texture is more pronounced parallel to the system plane and perpendicular to the transport direction ($y$), while the out-of-plane ($z$) spin texture is only significant in a small energy range in the conduction band right above the Fermi level. This unconventional spin-texture results from the combination of inversion asymmetry and QSH phase originated by the inverted band gap, this feature is also observed for IV-V half-functionalized QSH phases \cite{MeraAcosta2016a}.

\begin{figure}[htb]
	\centering
	\includegraphics[width=\linewidth]{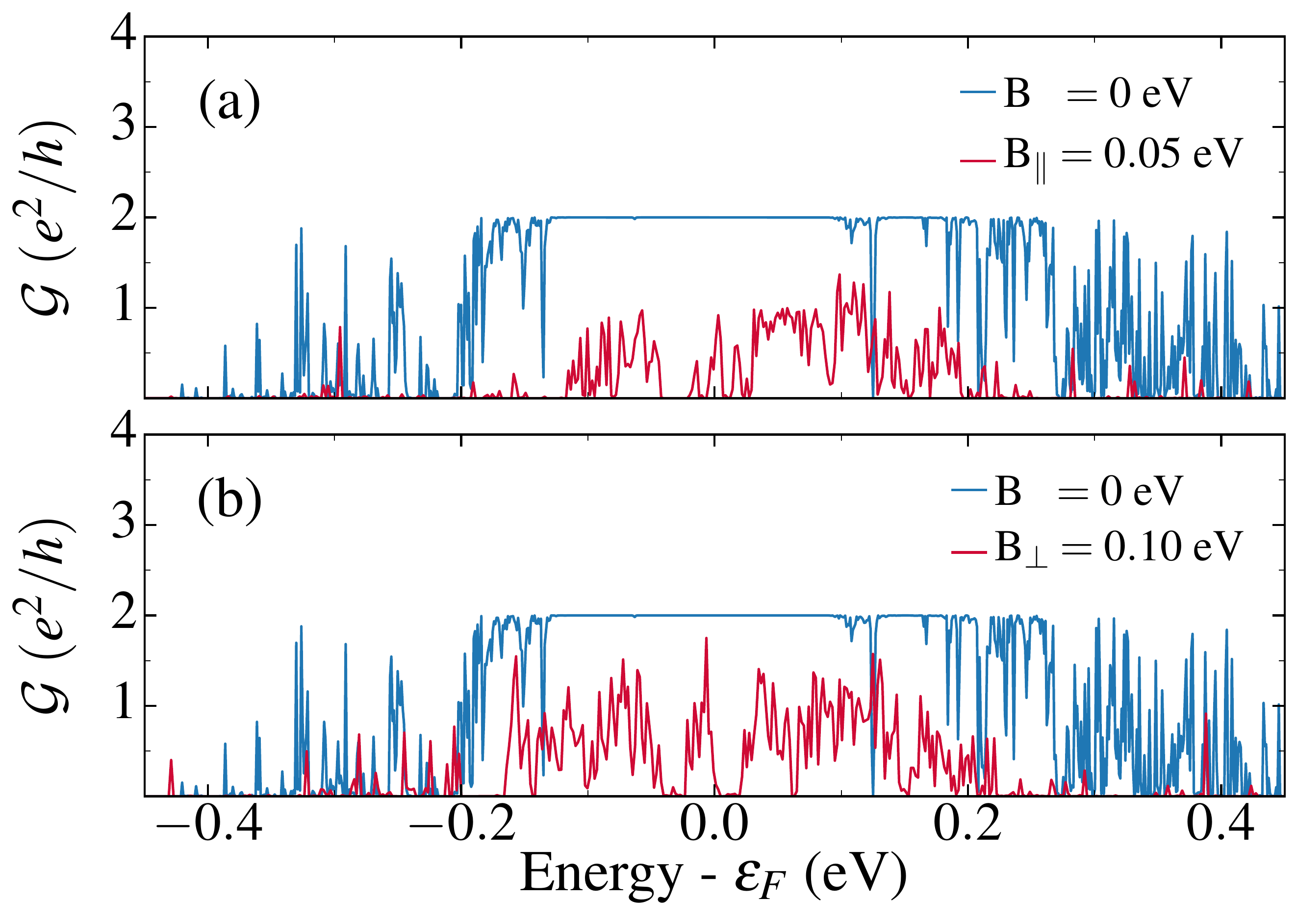}
	\caption{Electronic conductance with and without an on-site exchange field of the form $\delta H = \mathbf{B} \cdot \mathbf{\sigma}$ for (a) parallel ({\scriptsize$\parallel$}) and (b) perpendicular ({\scriptsize$\perp$}) directions.}
	\label{fig:trc_field}
\end{figure}

Figure \ref{fig:trc_field} shows the conductance of the amorphous topological insulator nanoribbons for parallel ($\delta H = B_\parallel \sigma_y$) and perpendicular ($\delta H = B_\perp \sigma_z$) field directions using the \SI{324}{\nano\metre} device setup of Fig. \ref{fig:trc_length}(d). The breaking of TRS erases the topological protection of the edge states, thus decreasing the system conductance independent of the field direction. We note that exchange fields in the out-of-plane direction (Fig. \ref{fig:trc_field}(b)) are less effective in suppressing ${\mathcal{G}}$ than in-plane ones (Fig. \ref{fig:trc_field}(a)). This is a consequence of the spin quantization axis in the nanoribbon, see Fig. \ref{fig:texture_field_bands}. The combination of realistic device length and exchange field induces localization effects, dramatically quenching the conductance as shown in Fig. \ref{fig:trc_field}.

\begin{figure}[htb]
	\centering
	\includegraphics[width=\linewidth]{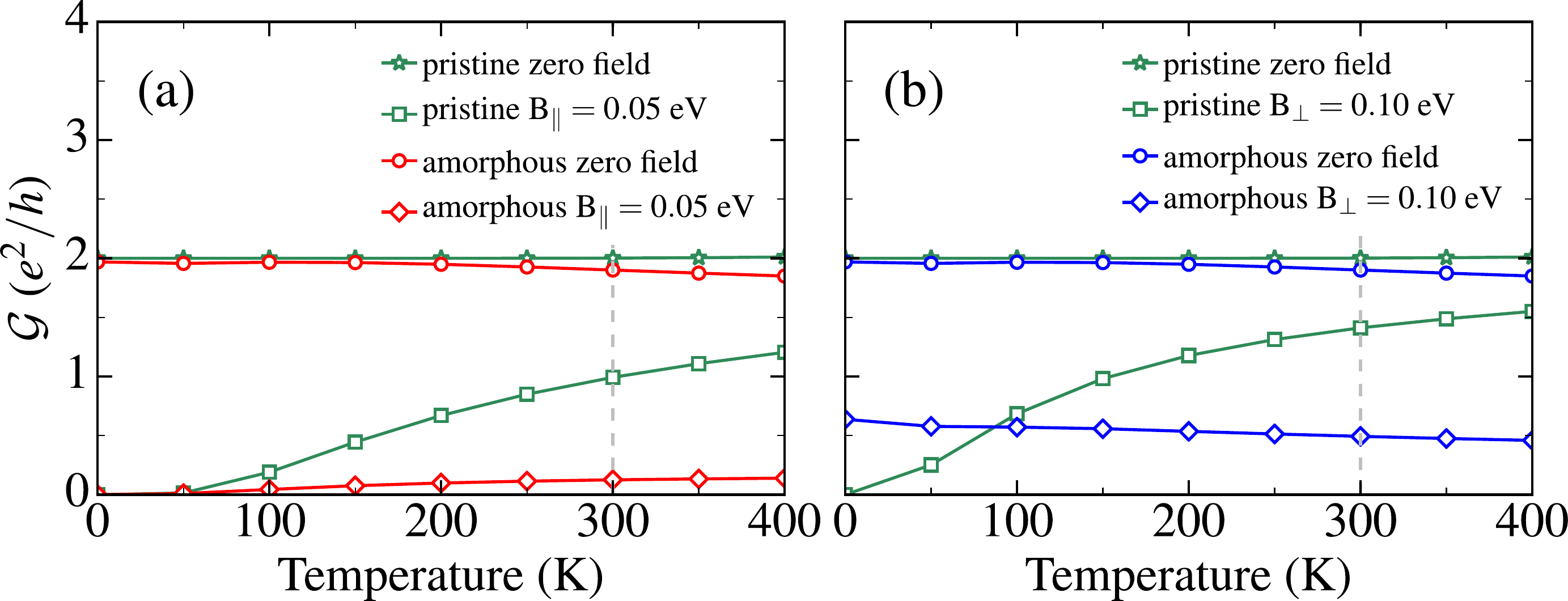}
	\caption{Electronic conductance at the system Fermi energy $\varepsilon_F$ for a randomly selected sequence of amorphous systems as a function of the temperature (T) for (a) parallel exchange field ($B_\parallel$) and (b) perpendicular exchange field ($B_\perp$).}
	\label{fig:differential_conductance}
\end{figure}

In Fig. \ref{fig:differential_conductance} we show the conductance at the system Fermi energy $\varepsilon_F$ as a function of temperature, Eq. \eqref{eq:cond_temp}, for a parallel, Fig. \ref{fig:differential_conductance}(a), and perpendicular, Fig. \ref{fig:differential_conductance}(b), on-site exchange field. We calculate the ratio between conductance for the system without and with the in-plane exchange field ($\mathcal{G}_{\rm off}/\mathcal{G}_{\rm on}$). For the pristine and amorphous devices, this ratio is $2.02$ and $15.10$ at \SI{300}{\kelvin}, respectively. The enhanced on/off ratio reveals that the amorphous TIs can aid the realization of QSHI-based devices at room temperature. This temperature effect is proportional to the band gap, and varies for different TIs \cite{Wu2018a,Focassio2020}. Realizing amorphous materials with topological properties is desirable to hinder the interference of bulk transport at finite temperature, screening only the transport response of topological edge states, as in Fig. \ref{fig:trc_length}(d) and Fig. \ref{fig:trc_field}(a). Hence, the amorphization may be used as a filtering technique for several topological systems.

\section{Summary and Conclusions}
\label{sec:conclusions}

We have studied the realization of amorphous topological insulator materials using state of the art computational techniques to study the structural, electronic and transport properties of structure that may be obtained by standard experimental techniques before annealing.

We use a random bond flip method combined with \textit{ab initio} calculations to generate flat bismuthene amorphous structures. The RDF shows that the obtained systems have short range and lack long range order. We characterize the structure by different statistical measures and find that the ring size distribution nicely captures the degree of amorphization of our structures. We find that amorphous bismuthene has a coordination that is significantly smaller than $z=3$, that is reflected in an average bond length smaller than the pristine one and, more interestingly, in the formation of holes.

We study the topological properties of the system by calculation the $\mathbb{Z}_2$ invariant and the spin Bott index. As expected, these invariants reveal the non-trivial topological band structure of the material. One of the main results of this study is that the amorphization tends to suppress the band gap, but does not closes it. We find that the survival of the QSH state through the amorphization process is associated with the SOC strength of the material and the size of the bulk band gap.

Interestingly, we also find that the amorphization-induced holes host localized states at the corresponding internal edges of the system. This, not only confirms the lack of bondings in these regions, but also gives rise to new non-trivial states that deserve further investigation.

Next, we investigate the Landauer conductance for devices with realistic lengths, up to \SI{324}{\nano\metre}. This is achieved by building nanoribbons composed of randomly selected sequences of the amorphous systems generated by \textit{ab initio} techniques. Using this setup we obtain that the topological helical edge states with conductance $2e^2/h$ are preserved inside the gap. For energies outside the topological gap region we find a strong suppression of the conductance, consistent with Anderson localization, a clear indication of a trivial insulator phase. Furthermore, the conductance is controllable by a simple exchange field that may be induced by a substrate or experimental probe.

Due to the robustness of such systems, amorphous H-bismuthene QSHIs may be the key to the experimental realization of QSHI-based devices at room temperature, not only displaying the gapless helical edge state but also hindering the bulk transport response.

\begin{acknowledgments}

This work is supported by FAPESP (Grants  
19/04527-0, 16/14011-2, 17/18139-6, and 17/02317-2), FAPERJ (Grants 
E-26/2020.882/2018 and E-26/010.101126/2018), and CNPq (Grant 
308801/2015-6). The authors acknowledge the Brazilian Nanotechnology National Laboratory (LNNano/CNPEM, Brazil) and the SDumont supercomputer at the Brazilian National Scientific Computing Laboratory (LNCC) for computational resources.

\end{acknowledgments}

\bibliographystyle{apsrev4-2}
\bibliography{references,amorphousTI}

\end{document}